\documentclass[aps,pra,onecolumn,superscriptaddress,nofootinbib]{revtex4-2}

\usepackage{amsmath,amssymb,braket,bm}
\usepackage{hyperref}

\begin{document}

\title{Detecting Hidden Nonlinear High Frequency Modes Beyond Fundamental Minimal Temporal Resolution Using Weak Measurements}

\author{Igor Prlina}
\affiliation{Institute of Physics, University of Belgrade, Pregrevica 118, 11080 Belgrade, Serbia}


\begin{abstract}
In this work we study whether nonlinear models of quantum mechanics can avoid detection by strong measurements, if the nonlinear effects only exist as high frequency modes. A potential physical process which could hide such modes would be amplitude level time averaging over fundamentally indistinguishable times. To that end, we have defined a temporal indistinguishability postulate which describes the averaging process, and handcrafted a nonlinear toy model well suited to avoid detection. We have demonstrated that even in such an ideal setup, tabletop weak postselected measurements can still detect nonlinear behavior and put constraints on nonlinear high frequency modes of evolution, even at frequencies beyond fundamental minimal temporal resolution.

\end{abstract}

\maketitle

\section{Introduction}

In standard quantum mechanics, it is assumed that the evolution of any state is linear. However, nonlinear modifications of quantum mechanics have been explored in many different contexts, as nonlinearity can provide a mechanism for different important phenomena, such as objective collapse of the wavefunction in the act of measurement, decoherence, and quantum chaos \cite{PenroseCollapse,GhirardiCollapse,DiosiCollapse,NonlinearCollapse,ObjectiveCollapseNew}. Nonlinear models face significant hurdles though. Practically, they are heavily constrained by experimental tests of linearity \cite{ObjectiveCollapseTest}, and theoretically they often lead to possibility of superluminal signaling. A question remains whether a type of nonlinear evolutions can exist in such a way to avoid these issues. A potential loophole arises from the fact that no measurement is ideal.

Quantum mechanics predicts the probability distribution in the act of measurement of a quantum system at some time $t$ via the Born rule. In realistic experiments, however, the time at which a measurement occurs cannot be specified with arbitrary precision. This limitation is usually considered to be instrumental. Due to a finite resolution of any time measuring device, we cannot ensure that we measure all members of the ensemble at the same instance of time. In effect, we are mixing different ensembles corresponding to different moments of measurement. As such, probabilities, expectation values, and projectors need to be averaged over a finite time interval corresponding to the precision of the clock. Under such averaging, nonlinear evolution with modes faster than the clock resolution can mimic mixed states \cite{QuickOscillations}.

Instead of the clock resolution, one can consider a fundamental temporal resolution, namely a time scale below which it is fundamentally impossible to distinguish measurement times (possibly as a consequence of some yet unknown quantum gravity effects or otherwise). Under such assumption, measurements at different moments of time within such an interval would be fundamentally indistinguishable. Using the standard quantum approach, amplitudes of fundamentally indistinguishable events need to be summed before probabilities are evaluated. This opens a potential loophole for nonlinearity. If nonlinear part of the evolution only has high frequency modes faster than the fundamental time scale, nonlinear effects would disappear under the amplitude summation, and these effects would appear to be fundamentally undetectable. While such nonlinear high frequency modes are unlikely to have practical application, given that they avoid detection, they could nevertheless have theoretical relevance. Such high frequency modes could, in principle, provide a hidden sector in which effects associated with nonlinearity are stored without appearing in ordinary measurements. For example, they may be relevant to questions involving superluminal signaling, apparent energy nonconservation, or information loss. Importantly, the nonlinear part of the wavefunction is not constrained to be small relative to the linear part.

In order to study the potential existence of high frequency nonlinear modes, one can try to apply the methodology of weak measurements \cite{Spin100, WeakMeasurement}. Namely, in addition to the standard strong measurements, one can also perform the so-called weak (postselected) measurements. In weak measurements, the device that interacts with the quantum system has quantum nature itself. The wavefunction of the device needs to be strongly delocalized, and the interaction strength with the system needs to be weak. Ensembles under weak measurements can undergo postselection. Postselection is a selective strong filtration measurement at the end of the experiment. If an ensemble member fails postselection, it is discarded from the ensemble together with its corresponding quantum device. The mean value of the surviving device pointer positions corresponds to the so-called weak value. Unlike transition amplitudes, weak values are not linear functionals of the state. This leads to non-trivial behavior under time averaging. In our previous work, we have demonstrated that time averaged weak measurements can distinguish pure states with nonlinear high frequency modes from mixed states without high frequency modes that the pure states with nonlinear high frequency modes mimic due to the averaging at the probability level inherent to the finite resolution of the clock \cite{QuickOscillations}. Postselected systems with linear nonunitary evolution have also been considered \cite{NonunitaryABL, DecoherenceWeak}. We have also shown that weak measurements can distinguish a finite resolution of the clock from a fundamental continuous nonunitary evolution \cite{ObjectiveWeak}.

The main goal of this paper is to demonstrate that even nonlinear effects that only introduce nonlinear high frequency modes beyond fundamental temporal resolution cannot evade detection if weak measurements are performed. We show that if there exists a fundamental temporal resolution, the weak value itself needs to be averaged. Given that weak values are nonlinear functionals of the state, the time averaging does not delete the nonlinear contribution. We study a handcrafted model of a branching nonlinear evolution which is constructed in such a way to easily avoid detection via strong measurements. We demonstrate that even for such a model, a ratio of the magnitudes of the linear and nonlinear part of the state can be detected. Thus, even the nonlinear high frequency modes beyond fundamental temporal resolution face strong experimental constraints.

The layout of the paper is as follows. In Section II we precisely define the temporal indistinguishability postulate. Section III considers the undetectability of nonlinear high energy modes using strong measurements. Next, in Section IV we demonstrate that weak values undergo temporal averaging due to the temporal indistinguishability postulate.  A simple example is given in Section V, where we explicitly show that, for a cubit, weak postselected measurements can detect nonlinear high frequency modes beyond fundamental temporal resolution. Finally, our conclusions are summarized in Section VI.

\section{Temporal indistinguishability postulate}

We will start by specifying the consequence on quantum kinematics stemming from the assumed existence of fundamental time uncertainty, which we call the \emph{temporal indistinguishability postulate (TIP):} 
before applying the Born rule to calculate probabilities, all transition amplitudes at sharply specified times,
\begin{equation}
    \mathcal{A}(t)=\braket{f|\psi(t)},
\end{equation} 
ought to be replaced by the effective time averaged amplitudes, 
\begin{equation}
    \mathcal{A}_{\rm eff}(t)
    =
    \int dt'\, w_\tau(t, t')\braket{f|\psi(t')},
\end{equation}
where $w_\tau(t, t')$ is the normalized temporal window associated with the fundamental indistinguishability scale $\tau$,
\begin{equation}
	w_{\tau}(t, t')=\frac{1}{\tau}H(t'-t+\frac{\tau}{2})H(t+\frac{\tau}{2}-t'), \label{w}
\end{equation}
and $H$ is the Heaviside step function.

The interpretation of the postulate is that there exists a fundamental time scale $\tau$ such that measurement events separated by less than $\tau$ cannot be distinguished even in principle. Amplitudes corresponding to transitions occurring at indistinguishable times must be summed prior to forming probabilities.
Note that this is distinct from instrumental averaging due to the finite resolution of the clock, where one averages probabilities or projectors:
\begin{equation}
    P_{\rm instr}
    =
    \int dt'\, w(t, t')\,|\braket{f|\psi(t')}|^2.
\end{equation}
The distinction is essential, since the two averaging approaches lead to different experimental signals. Note that since the postulate holds for transitions into any state, the postulate is equivalent to substituting time dependent states with effective averaged states directly. Finally, while we take the normalized temporal window to be described by \eqref{w} throughout this paper, we do so merely for the sake of simplicity, and its precise form can be modified if needed. 

\section{Undetectability of Nonlinear High Energy Modes}

Due to TIP, some potential nonlinear effects can avoid detection due to averaging at the amplitude level. In this Section we will present a nonlinear toy model handcrafted to avoid experimental constraints, as follows. In linear quantum mechanics, two vectors differing by a nonzero complex factor represent the same physical state:
\begin{equation}
    \ket{\psi} \sim C\ket{\psi}.
\end{equation}
Linear homogeneous evolution preserves this equivalence,
\begin{equation}
    E(t,C\ket{\psi}) = C E(t,\ket{\psi}),
\end{equation}
while nonlinear evolution can be nonhomogeneous,
\begin{equation}
    E(t,C\ket{\psi}) \neq C E(t,\ket{\psi}).
\end{equation}
Thus, if evolution is allowed to be nonlinear, two vectors differing only in norm can evolve differently. This allows for a model of nonlinear evolution described by two distinct branches, the first being the standard evolution of linear quantum mechanics, while the second contains additional high frequency modes.
Therefore, we will consider the following nonlinear model:
\begin{equation}
E(t-t_0)\ket{\Psi(t_0)}
=
\begin{cases}
e^{-iH_q(t-t_0)}\ket{\Psi(t_0)},&
\braket{\Psi(t_0)|\Psi(t_0)}\geq N_c^2,\\[0.5em]
e^{-iH_s(t-t_0)}\ket{\Psi(t_0)},&
\braket{\Psi(t_0)|\Psi(t_0)}<N_c^2.
\end{cases}
\label{eq:branching}
\end{equation}
Here, the modified Hamiltonian $H_q$ is defined by
\begin{equation}
   H_q=H_s+H_{hf},
\end{equation}
where $H_s$ is the standard linear Hamiltonian, and $H_{hf}$ is a linear Hermitian operator which contains only two nonempty sets of eigenvalues: high frequency modes at time scales lower than the fundamental indistinguishability scale, and zeros.
The parameter $N_c$ is the critical norm and can be arbitrarily large in principle.  Note that the evolution is both linear and unitary for any fixed value of the norm of the wavefunction (albeit different norms correspond to different Hamiltonians).
On the other hand, the full evolution is nonlinear, since it explicitly depends on the value of the norm, which is quadratic in the wavefunction, violating the superposition principle.

There are two properties of the operator $H_{hf}$ that we need to elaborate on. First, we said that all its nonzero eigenvalues are "high frequency". In nonrelativistic physics, energy is defined up to an arbitrary constant, that can be very large in principle. As such, a high frequency Hamiltonian can become a low frequency Hamiltonian and vice versa, via an energy redefinition. However, in General Theory of Relativity, energy is gravitational charge and the energy constant is fixed by the gravitational interaction. Therefore, throughout this paper, energies are defined such that the energy of a noninteracting particle at rest is given by $E=mc^2$. Second, if all components of a state evolve under high frequencies, after applying TIP the effective state would become zero. A zero state does not have a valid interpretation, and as such we require  $H_{hf}$ to contain at least one null eigenvalue.

There are two important time scales to consider, namely the resolution of the clock $ \Delta t_{\rm exp}$, and the fundamental time indistinguishability scale $\tau$. If the frequencies in $H_{hf}$ are lower than the frequency corresponding to $ \Delta t_{\rm exp}$, the nonlinearity is fully detectable and constrained by the standard tests of nonlinearity. Thus, we will not consider this frequency range. The two frequency regimes of interest for our paper are when the frequencies in $H_{hf}$ are in between the frequencies corresponding to $ \Delta t_{\rm exp}$ and $\tau$, or when the frequencies in $H_{hf}$ are above the frequency corresponding to $\tau$.  We will illustrate the behavior in these regimes on a two-level state,
\begin{equation}
    \ket{\psi(t)}
    =
   N_q \left(a\ket{+}
    +
    b e^{-i\Omega t}\ket{-}\right),
    \label{eq:state}
\end{equation}
where $N_q$ is the norm of the state above the threshold norm $N_c$, and $a$ and $b$ are complex parameters that satisfy
\begin{equation}
   |a|^2+|b|^2=1.
    \label{eq:normq}
\end{equation}
The evolution of this state corresponds to the choice $H_s=0$ (the evolution of the slowly evolving branch is of negligible influence on our considerations, so we are free to use the approximation that the slowly evolving branch does not change in time at all) and $H_{hf}=\mathrm{diag}\{0, \hbar \Omega\}$ in the $\{\ket{+}, \ket{-}\}$ basis.

\subsection{Instrumentally undetectable but fundamentally observable oscillations}

If the oscillation is fast relative to experimental time resolution but slow relative to the fundamental scale,
\begin{equation}
   \tau\ll \Omega^{-1}\ll \Delta t_{\rm exp} ,
\end{equation}
then the direct state averaging due to TIP does not lead to observable changes. Namely, since the state is a very slowly changing function in time, it is almost invariant to averaging over a short interval. In the temporal indistinguishability interval, $|\Omega(t'-t)|\ll 1$, the state \eqref{eq:state} can be rewritten as
\begin{equation}
    \ket{\psi(t')}
    =
   N_q \left(a\ket{+}
    +
    b e^{-i\Omega (t'-t)}e^{-i\Omega t}\ket{-}\right).
    \label{eq:staterewritten}
\end{equation}
After power expanding the first exponent up to linear terms, we obtain
\begin{equation}
    \ket{\psi(t')}
    =
  \ket{\psi(t)}+\Delta\ket{\psi(t')},
    \label{eq:statepluscorrection}
\end{equation}
where
\begin{equation}
    \Delta\ket{\psi(t')}
    =
   -i N_q b e^{-i\Omega t}\Omega (t'-t)\ket{-}.
    \label{eq:statecorrection}
\end{equation}
The term $ \Delta\ket{\psi(t')}$ is the only part of the state that depends on $t'$ and as such, the TIP averaging only affects this summand. After applying TIP,
\begin{equation}
    \Delta\ket{\psi(t)}_{eff}=\frac{1}{\tau}\int_{t-\tau/2}^{t+\tau/2} \Delta\ket{\psi(t')} \,dt'
    =
   -i \frac{N_q \Omega b}{\tau} e^{-i\Omega t}\int_{-\tau/2}^{\tau/2} (t'-t) \,d(t'-t)\ket{-}=0.
    \label{eq:statecorrectionaveraged}
\end{equation}
Therefore, corrections to the state due to the application of TIP only start occurring at quadratic order in $\Omega \tau$.

On the other hand, the finite resolution of the clock used in the experiment, $ \Delta t_{\rm exp}$, does lead to a significant effect in this regime, since $\Omega  \Delta t_{\rm exp}\gg 1$. The experimental protocol corresponds to averaging projectors or probabilities instead of amplitudes. The off-diagonal terms in
\begin{equation}
    \ket{\psi(t)}\bra{\psi(t)}
\end{equation}
average out to zero, and the effective state becomes
\begin{equation}
    \rho_{\rm eff}
    =
    |a|^2\ket{+}\bra{+}
    +
    |b|^2\ket{-}\bra{-}.
\end{equation}
In this regime, states undergoing evolution with nonlinear high frequency modes mimic mixed states under strong nonpostselected measurements. This regime was studied in detail in \cite{QuickOscillations}, and an experimental protocol for its detection based on weak postselected measurements was proposed.

\subsection{Fundamentally fast oscillations}

The last regime is the main interest of this paper. If the additional frequencies of the nonlinear branch are at time scales below the fundamental indistinguishability scale,
\begin{equation}
    \Omega^{-1} \gg \tau,
\end{equation}
then TIP has nontrivial consequences. The nonlinear high frequency mode is averaged out at the amplitude level,
\begin{equation}
    \int dt'\, w_\tau(t,t')\, e^{-i\Omega t'}\approx 0,
\label{eq:averegedout}
\end{equation}
since the integral of the phase over a period is zero, the temporal window contains many periods, and the edge part of the temporal window that does not complete a full period is suppressed by the division by the temporal scale $\tau$ due to averaging.
Therefore, TIP requires that the quickly oscillating state
\begin{equation}
 \ket{\psi(t)}
    =
   N_q \left(a\ket{+}
    +
    b e^{-i\Omega t}\ket{-}\right),
\end{equation}
needs to be replaced by the effective state
\begin{equation}
    \ket{\psi(t)}_{eff}
    =
N_q a\ket{+}
\end{equation}
in all expressions.

The component $b\ket{-}$ becomes fundamentally unobservable via strong nonpostselected measurements. They cannot distinguish the state
\begin{equation}
    \ket{\psi_0 (t)}=Na\ket{+}
\end{equation}
of any norm from the supercritical norm state
\begin{equation}
    \ket{\psi(t)}
    =N_q \left(
    a\ket{+}
    +
    b e^{-i\Omega t}\ket{-}\right),
    \qquad \tau\gg\Omega^{-1}.
\end{equation}
However, we will show that weak postselected measurements can still distinguish these states, despite the fundamental averaging imposed by TIP.

\section{Weak measurements under temporal averaging}

Strong measurements describe outcomes of strong interactions between a quantum system and a classical (well localized) device. Weak postselected measurements describe outcomes of weak interactions between a quantum system and a quantum (delocalized) device, which survive postselection. Only the devices which correspond to ensemble members that have satisfied the postselection test are taken into account. From the standard theory of weak measurements \cite{Spin100, WeakMeasurement}, we know that if the initial device state is taken to be a very delocalized Gaussian, 
\begin{equation}
\Phi_0(q)
=
\frac{1}{(2\pi\sigma^2)^{1/4}}
\exp\left(-\frac{q^2}{4\sigma^2}\right),
\end{equation}
after a weak von Neumann interaction of an observable $A$ and the device momentum $P$ at time $t$, followed by postselection on the state $\ket{f}$, the pointer amplitude is approximately shifted by the so-called weak value,
\begin{equation}
    A_w(t)
    =
    \frac{\bra{f}A\ket{\psi(t)}}{\braket{f|\psi(t)}},
\end{equation}
and the pointer state becomes
\begin{equation}
\Phi(q,t)
=
\frac{1}{(2\pi\sigma^2)^{1/4}}
\exp\left[
-\frac{(q-gA_w(t))^2}{4\sigma^2}
\right],
\end{equation}
where $g$ is the weak interaction strength.

Note that we stated that the device needs to be very delocalized and the interaction strength needs to be very small,
\begin{equation}
g|A_w(t)| \ll \sigma.
\end{equation}
In practice, this means that the expression is only accurate to linear order in the weak value. Thus, the exponent can be expanded to a power series to linear order,
\begin{align}
\exp\left[
-\frac{(q-gA_w(t))^2}{4\sigma^2}
\right]
&\approx
\exp\left(-\frac{q^2}{4\sigma^2}\right)
\left[
1+\frac{gq}{2\sigma^2}A_w(t)
\right].
\end{align}
According to TIP, amplitudes corresponding to different interaction times must be summed before probabilities are formed. Thus the effective pointer amplitude is
\begin{equation}
\Phi_{\rm eff}(q,t)
=
\int dt'\,w_{\tau}(t,t')\,\Phi(q,t')
\end{equation}
At linear order, the averaging gives
\begin{align}
\Phi_{\rm eff}(q)
&\approx
\frac{1}{(2\pi\sigma^2)^{1/4}}
\exp\left(-\frac{q^2}{4\sigma^2}\right)
\left[
1+\frac{gq}{2\sigma^2}
\int dt'\,w_{\tau}(t,t')A_w(t')
\right].
\end{align}
Let us define the time-averaged weak value as
\begin{equation}
\bar{A}_w
=
\int dt'\,w_{\tau}(t,t')A_w(t').
\end{equation}
The effect of time averaging is that the components in the Fourier transform of $A_w$ with frequencies beyond the fundamental temporal indistinguishability scale get truncated, while components with frequencies lower than the same scale remain unchanged at linear order in $\omega \tau$. The derivation is the same as in \eqref{eq:averegedout} and \eqref{eq:statecorrectionaveraged}, although note that the Fourier transform of $A_w$ does not correspond to the Fourier transform of the state $\ket{\psi(t)}$ itself.

After averaging,
\begin{equation}
\Phi_{\rm eff}(q)
\approx
\frac{1}{(2\pi\sigma^2)^{1/4}}
\exp\left(-\frac{q^2}{4\sigma^2}\right)
\left[
1+\frac{gq}{2\sigma^2}\bar{A}_w
\right].
\end{equation}
This is precisely the first-order power series expansion of a Gaussian shifted by $g\bar A_w$,
\begin{equation}
\Phi_{\mathrm{eff}}(q,t)
=
\frac{1}{(2\pi\sigma^2)^{1/4}}
\exp\left[
-\frac{(q-g\bar{A}_w(t))^2}{4\sigma^2}
\right].
\end{equation}
Therefore, when TIP is applied, in the weak interaction strength limit, the pointer amplitude is shifted by the time-averaged weak value. Consequently, the mean shift of the device, which is the directly observable quantity, is the time averaged weak value.

It is important to stress two things. First, the need to time average the weak value is not an additional assumption. It is a direct consequence of TIP. Second, the presented derivation is not approximate, since the weak values are only defined in the weak interaction strength limit up to linear terms. The only remaining question is whether the weak value formalism is applicable in the case of nonlinear evolution which we are considering in this work. The answer is yes, and the applicability has been shown in \cite{QuickOscillations}. We will not repeat the full proof here, but instead only note that the proof is based on two facts. First, while the full evolution is nonlinear, in each individual branch it is unitary. Second, the weak value expression is invariant to a combination of taking an adjoint and time reversal together. These two facts are enough to prove the applicability of the weak value formalism in the considered case.

\section{Two-level example}

As we have shown in the previous section, a postselected weak measurement protocol directly observes the time averaged weak value, assuming that TIP holds. Since the weak value is not a linear function in quantum state, it enables detection of high frequency nonlinear modes that are undetectable otherwise. We will demonstrate how nonlinear high frequency modes can be detected in the case of a two-level quantum system.

We will consider the quantum system state
\begin{equation}
    \ket{\psi(t)}
    =N_q\left(
    a\ket{+}
    +
    b e^{-i\Omega t}\ket{-}\right),
\end{equation}
with $N_q>N_c$, and the postselected state
\begin{equation}
    \ket{f}
    =N_s \left(
    \alpha\ket{+}
    +
    \beta\ket{-} \right),
\end{equation}
with $N_s<N_c$.
We take the observable to be the spin/polarization operator,
\begin{equation}
    A=\sigma_z
    =
    \ket{+}\bra{+}
    -
    \ket{-}\bra{-}.
\end{equation}
The denominator of the weak value becomes
\begin{equation}
    \braket{f|\psi(t)}
    =N_q N_c \left(
    \alpha^* a
    +
    \beta^* b e^{-i\Omega t} \right),
\end{equation}
while the numerator is obtained from
\begin{equation}
    \sigma_z\ket{\psi(t)}
    = N_q \left(
    a\ket{+}
    -
    b e^{-i\Omega t}\ket{-} \right),
\end{equation}
leading to
\begin{equation}
    \bra{f}\sigma_z\ket{\psi(t)}
    = N_q N_c \left(
    \alpha^* a
    -
    \beta^* b e^{-i\Omega t} \right).
\end{equation}
Thus the weak value is
\begin{equation}
    (\sigma_z)_w(t)
    =
    \frac{\alpha^* a - \beta^* b e^{-i\Omega t}}
         {\alpha^* a + \beta^* b e^{-i\Omega t}}.
\end{equation}
Note that the weak value does not depend on the explicit norms of the preselected and the postselected state. The expression is valid as long as the preselected state belongs to the supercritical norm branch of the evolution, while the postselected state belongs to the subcritical norm branch. 
Let us introduce the notation
\begin{equation}
    D_0 = \alpha^* a,
    \qquad
    D_1 = \beta^* b,
\end{equation}
and rewrite the weak value expression,
\begin{equation}
    (\sigma_z)_w(t)
    =
    \frac{D_0 - D_1 e^{-i\Omega t}}
         {D_0 + D_1 e^{-i\Omega t}}.
\end{equation}

Now we need to find the time-averaged weak value. As shown in the previous section, due to TIP the directly observable weak value takes the form
\begin{equation}  
\overline{(\sigma_z)_w}
    =
    \frac{1}{\tau}
  \int_{t-\frac{\tau}{2}}^{t+\frac{\tau}{2}}
    dt'\,
     \frac{D_0 - D_1 e^{-i\Omega t'}}
         {D_0 + D_1 e^{-i\Omega t'}}.
    \label{eq:weakTIP}
\end{equation}
The integrand is a periodic function with the period $T=\frac{2\pi}{\Omega}$. Since $\Omega \tau \gg 1$, the time interval $\tau$ will contain $N \gg 1$ full periods and one incomplete oscillation,
\begin{equation}
   \tau=N T+\Delta \tau.
\end{equation}
The time-averaged weak value becomes
\begin{equation}  
\overline{(\sigma_z)_w}
    =
    \frac{N}{\tau}
  \int_{0}^{T}
    dt'\,
     \frac{D_0 - D_1 e^{-i\Omega t'}}
         {D_0 + D_1 e^{-i\Omega t'}}
+
 \frac{1}{\tau}
  \int_{t+\frac{\tau}{2}-\Delta \tau}^{t+\frac{\tau}{2}}
    dt'\,
     \frac{D_0 - D_1 e^{-i\Omega t'}}
         {D_0 + D_1 e^{-i\Omega t'}}.
    \label{eq:weakTIPsplit}
\end{equation}
In zeroth order of power expansion in $\frac{\Delta \tau}{\tau}$, the second integral vanishes and the time averaged weak value becomes
\begin{equation}  
\overline{(\sigma_z)_w}
    =
    \frac{1}{T}
  \int_{0}^{T}
    dt'\,
     \frac{D_0 - D_1 e^{-i\Omega t'}}
         {D_0 + D_1 e^{-i\Omega t'}}.
    \label{eq:weakTIPapprox}
\end{equation}
We can integrate over the phase instead of time, 
\begin{equation}
    \overline{(\sigma_z)_w}
    =
    \frac{1}{2\pi}
    \int_0^{2\pi}
    d\theta\,
    \frac{D_0 - D_1 e^{-i\theta}}
         {D_0 + D_1 e^{-i\theta}}.
    \label{eq:phaseintegral}
\end{equation}
We can switch to a complex variable $z = e^{-i\theta}$, so that $d\theta = -dz/(iz)$. The integral then becomes a contour integral over the unit circle,
\begin{equation}
    \overline{(\sigma_z)_w}
    =
    \frac{1}{2\pi i}
    \oint_{|z|=1}
    dz\,
    \frac{D_0 - D_1 z}{z(D_0 + D_1 z)}.
    \label{eq:contour}
\end{equation}
To solve this integral, we can evaluate its residues. The integrand has two simple poles,
\begin{equation}
    z = 0,
    \qquad
    z_* = -\frac{D_0}{D_1}.
\end{equation}
The residue at $z=0$ is
\begin{equation}
    \operatorname{Res}_{z=0} = 1,
\end{equation}
while the residue at $z=z_*$ is
\begin{equation}
    \operatorname{Res}_{z=z_*} = -2.
\end{equation}
The integral can have different values depending on which poles are inside the unit circle.  If $|D_0| > |D_1|$, then $|z_*| > 1$ and only the pole at $z=0$ contributes. If $|D_1| > |D_0|$, then both poles contribute.
Thus, we obtain the time averaged weak value,
\begin{equation}
    \overline{(\sigma_z)_w}
    =
    \begin{cases}
    +1,& |D_0| > |D_1|,\\[0.4em]
    -1,& |D_1| > |D_0|.
    \end{cases}
\label{eq:signflip}
\end{equation}
By returning to the original parameters,
\begin{equation}
    |D_0| = |\alpha||a|,
    \qquad
    |D_1| = |\beta||b|,
\end{equation}
we can see that the transition occurs at
\begin{equation}
    |\alpha||a| = |\beta||b|.
\end{equation}
Operationally, this corresponds to a sharp transition in the measured weak value as the postselection state is varied. Thus, the detectable signal of high frequency nonlinear modes would be a discontinuous sign flip in the measured  (time-averaged) weak value. 

If there are no high frequency nonlinear modes, then $b=0$ and
\begin{equation}
    \ket{\psi}=N_q a\ket{+}.
\end{equation}
In that case, the weak value is easy to find. The denominator and the numerator in the weak value expression become
\begin{equation}
    \braket{f|\psi}=N_q N_s\alpha^*a,
\end{equation}
and
\begin{equation}
    \bra{f}\sigma_z\ket{\psi}=N_q N_s \alpha^*a,
\end{equation}
respectively. The weak value becomes
\begin{equation}
    (\sigma_z)_w=+1
\end{equation}
for every possible postselection condition with $\alpha\neq0$.
On the other hand, if the high frequency nonlinear mode exists but is hidden by fundamental temporal indistinguishability, then the sign flip in Eq.~\eqref{eq:signflip} occurs. The transition condition depends on relative magnitude of the high frequency mode,
\begin{equation}
    |\alpha||a|=|\beta||b|.
\end{equation}
Thus, the weak measurement protocol does not only detect the presence of high frequency nonlinear modes. It can do more, namely it can measure the relative contribution of the high frequency mode by tuning the postselection state until the sign flip occurs. The transition occurs when the following relation is satisfied,
\begin{equation}
    \frac{|b|}{|a|}
    =
    \left.
    \frac{|\alpha|}{|\beta|}
    \right|_{\rm flip}.
\end{equation}
Therefore, the magnitude of the high frequency nonlinear mode invisible to all strong measurements can be operationally determined by weak postselected measurements.

To the best of our knowledge, such sudden sign flips have not been experimentally detected in weak measurements of polarization or spin in a given direction. Thus, even for the handcrafted toy model of nonlinear evolution designed to avoid detection, weak measurements put constraints on nonlinearity, even at frequencies beyond the fundamental temporal indistinguishability scale. Of course, an unexpected detection of such a sign flip would indicate that a modification of standard quantum mechanics would likely be necessary.

\section{Concluding Remarks}

In this work, we have considered whether nonlinear models of quantum mechanics can avoid detection by the effects on nonlinearity only existing in the high frequency regime fundamentally inaccessible to strong measurement. To that end, we have defined the temporal indistinguishability postulate, and we have handcrafted a nonlinear toy model so that it optimally avoids detection via strong measurements due to time averaging. We have demonstrated that even with such favorable assumptions, nonlinearity can still be constrained by weak postselected measurements.

The main application of our results is in constraining potential high energy physics models. Namely, most models of high energy physics involve some sort of a fundamental cut-off or coarse-graining (for example, the string size in string theory or cell size in loop quantum gravity). This work clearly demonstrates that it is not sufficient to test whether averaging over the fundamentally unreachable degrees of freedom reproduces low energy physics correctly on the level of strong measurements. It is also necessary to see if the averaging is consistent on the level of weak nonpostselected measurements as well. We have clearly shown that there exist at least some high energy theories that modify physics in such a way that low energy strong measurements give the same results as standard low energy quantum mechanics as expected, but for which there remains a detectable difference at the level of weak measurements. Our results are also relevant for the theoretical study of nonlinear models. Nonlinear models are known to potentially lead to superluminal signaling. One approach to alleviate this would be to have all the superluminal information transfer be present in the high frequency regime, where such signals cannot be detected in principle. In such a way, these theories could be compatible with the special theory of relativity and causality. However, in our work, we have demonstrated that such superluminal signals can still be detected by weak postselected measurements, at least for some of those models. Finally, we can conjecture that tabletop weak postselected measurements can be used to constrain at least some models of high energy physics. Weak postselected measurements are rarely considered as a potential probe of high energy physics, and we believe that they can be a useful tool for high energy experiments.

There are multiple possible avenues for future work. A systematic study of weak measurements of more complex nonlinear models, especially different objective collapse candidates, can be of interest. Additionally, the behavior of specific high energy models under weak measurements can be researched in detail. 

\section{Acknowledgments}

The author would like to thank Marko Vojinovi\'{c} and Nikola Paunkovi\'{c} for useful discussions. Research supported by the Ministry of Science, Technological Development and Innovations (MNTRI) of the Republic of Serbia.

\bibliographystyle{apsrev4-2}

\begin{thebibliography}{99}

\bibitem{PenroseCollapse}
R.~Penrose,
\emph{On gravity's role in quantum state reduction},
Gen. Relativ. Gravit. \textbf{28} (1996), 581--600.

\bibitem{GhirardiCollapse}
G.~C.~Ghirardi, P.~Pearle, and A.~Rimini,
\emph{Markov processes in Hilbert space and continuous spontaneous localization of systems of identical particles},
Phys. Rev. A \textbf{42} (1990), 78.

\bibitem{DiosiCollapse}
L.~Di\'osi,
\emph{A universal master equation for the gravitational violation of quantum mechanics},
Phys. Lett. A \textbf{120} (1987), 377--381.

\bibitem{NonlinearCollapse}
L.~Mertens, M.~Wesseling, N.~Vercauteren, A.~Corrales-Salazar, and J.~van Wezel,
\emph{Inconsistency of linear dynamics and Born's rule},
Phys. Rev. A \textbf{104} (2021), 052224.

\bibitem{ObjectiveCollapseNew}
L.~Mertens, M.~Wesseling, and J.~van Wezel,
\emph{Stochastic field dynamics in models of spontaneous unitarity violation},
SciPost Phys. Core \textbf{7} (2024), 012.

\bibitem{ObjectiveCollapseTest}
Vinante, A., R. Mezzena, P. Falferi, M. Carlesso, and A. Bassi. \emph{Improved Noninterferometric Test of Collapse Models Using Ultracold Cantilevers}, Physical Review Letters 119, no. 11 (2017): 110401.

\bibitem{QuickOscillations}
I.~Prlina,
\emph{If mixed states are secretly quickly oscillating pure states, weak measurements can detect it},
J. Phys. A \textbf{58} (2025), 475302.

\bibitem{Spin100}
Y.~Aharonov, D.~Z.~Albert, and L.~Vaidman,
\emph{How the result of a measurement of a component of the spin of a spin-$1/2$ particle can turn out to be 100},
Phys. Rev. Lett. \textbf{60} (1988), 1351.

\bibitem{WeakMeasurement}
Y.~Aharonov and L.~Vaidman,
\emph{Properties of a quantum system during the time interval between two measurements},
Phys. Rev. A \textbf{41} (1990), 11.

\bibitem{NonunitaryABL}
I.~P.~Prlina and N.~N.~Nedeljkovi\'{c},
\emph{Time-symmetrized description of nonunitary time asymmetric quantum evolution},
J. Phys. A \textbf{49} (2015), 035301.

\bibitem{DecoherenceWeak}
L.~B.~Ferraz, J.~Martin, and Y.~Caudano,
\emph{On the relevance of weak measurements in dissipative quantum systems},
Quantum Sci. Technol. \textbf{9} (2024), 035029.

\bibitem{ObjectiveWeak}
I.~Prlina and M.~Živković,
\emph{If Quantum Measurements Are Secretly Continuous Nonunitary Processes, Weak Measurements Can Detect It},
arXiv:2603.18787.




\end{thebibliography}

\end{document}